# Internal energy in compressible Poiseuille flow

Konrad Gizynski[1], Karol Makuch[1], Jan Paczesny[1], Yirui Zhang[1], Anna Maciołek[1,2]

and Robert Holyst[1*]

[1]Institute of Physical Chemistry, Polish Academy of Sciences
Kasprzaka 44/52, 01-224 Warsaw, Poland

[2]Max-Planck-Institut für Intelligente Systeme,
Heisenbergstr. 3, D-70569 Stuttgart, Germany

**ABSTRACT**

We analyse a compressible Poiseuille flow of ideal gas in a plane channel. We provide the form of internal energy $U$ for a non-equilibrium stationary state (NESS) that includes viscous dissipation and pressure work. We demonstrate that U depends strongly on the ratio $\Delta p/p_0$, where $\Delta p$ is the pressure difference between inlet and outlet and $p_0$ is the outlet's pressure. In addition, $U$ depends on two other variables: the channel aspect ratio and the parameter equivalent to Reynolds number. The stored internal energy, $\Delta U=U-U_0$, is small compared to the internal energy $U_0$ of the equilibrium state (ES) for a moderate range of values of $\Delta p/p_0$. However, $\Delta U$ can become large for big $\Delta p$ or close to vacuum conditions at the outlet ($p_0 \approx 0$ Pa).

**I. INTRODUCTION**

Poiseuille flow has been studied extensively for years [1] as it is ubiquitous in nature and essential from the application point of view [2–6]. However, most of these studies focus on velocities or pressure distribution in the flow while ignoring changes in internal energy compared to the equilibrium state, e.g. due to inhomogeneous temperature. Such approximations are especially common in the studies of the compressible version of the flow as it is described by non-linear equations and difficult to calculate analytically [7–14]. Frequently, either the problem is simplified to isothermal case [7,9–13] or the analysis is limited to the determination of approximate temperature profiles [8,14].

However, systems under flow are essentially non-equilibrium systems, and their macroscopic behaviour can be closely related to thermodynamic quantities. For example effects such as heating due to viscous dissipation can become important in micro-channel flows [15,16]. Furthermore, part of the energy supplied externally to sustain the flow can be stored inside the system in the form of internal energy. In our opinion, this energy excess $\Delta U$ can be essential in explaining the macroscopic behaviour of the system even though it is small compared to the total internal energy in the equilibrium state (ES) [17]. Thus, the knowledge of the exact expression for the internal energy and its relevant parameters can be crucial in explaining phenomena so complicated as turbulence.

The current paper is a continuation of our studies on energy stored in stationary states of non-equilibrium systems [17–19]. In our previous paper, we analysed a closed, non-equilibrium system that exchanges energy with the environment only in the form of heat [17]. We found that its internal energy has the following structure:

$$U = U_0 f \qquad (1)$$

where $U_0$ is the internal energy in equilibrium, and $f$ is a dimensionless function. In a further study, we focused on Poiseuille flow of incompressible ideal gas between two parallel plates where only viscous dissipation was responsible for internal energy changes [18]. We found that for such system $U$ can be expressed by the following equation:

$$U = U_0 \left(1 + \frac{1}{240} \frac{\Delta p^2 L_y^4}{\mu k T_0 L_x^2}\right), \qquad (2)$$

where $\Delta p$ is the pressure difference between the inlet and the outlet, $L_x$ and $L_y$ correspond to the length and the width of the channel, $\mu$ and $k$ represent dynamic viscosity and thermal conductivity of the gas, respectively, while $T_0$ denotes the temperature of the wall used as a boundary condition for NESS. Here we analyse if a similar form can be applied to describe internal energy in the compressible flow of ideal gas where both viscous dissipation and pressure work are present.

## II. MODEL AND METHODOLOGY

In the present paper we numerically study compressible Poiseuille flow of an ideal gas through a channel in two spatial dimensions. We focus on the stationary state. For monoatomic ideal gas, the three conservation laws for the mass, the momentum and the energy governing the behaviour of the system in the stationary state, are given by [20,21]:

$$\nabla(\rho \vec{u}) = 0, \tag{3}$$

$$\rho \vec{u} \cdot \nabla \vec{u} = -\nabla p + \nabla \cdot \left[ \mu \left( \nabla \vec{u} + (\nabla \vec{u})^T - \frac{2}{3} (\nabla \cdot \vec{u}) \mathbf{I} \right) \right], \tag{4}$$

$$\rho c_p \vec{u} \cdot \nabla T = \nabla \cdot (k \nabla T) + \vec{u} \cdot \nabla p + \Phi, \tag{5}$$

where $\rho$ is the density, $\vec{u}$ is the velocity field vector, $p$ is the pressure, $\mathbf{I}$ is the identity 3-dimensional matrix, $c_p$ is the specific heat capacity at constant pressure and $T$ is the temperature. For the compressible fluid, the viscous dissipation function $\Phi$ has the following form:

$$\Phi = \left[ -\frac{2}{3} \mu \nabla \cdot \vec{u} \mathbf{I} + \mu (\nabla \vec{u} + (\nabla \vec{u})^T) \right] : \nabla \vec{u}. \tag{6}$$

Additionally, there are two equations of state describing ideal gas:

$$RT\rho = Mp, \tag{7}$$

$$U = c_v nMT, \tag{8}$$

where $R$ is the gas constant, $M$ is the mean molar mass, $c_v$ corresponds to specific heat capacity at constant volume and $n$ is the number of moles. The equilibrium values of thermodynamic quantities are denoted by $p_0$, $T_0$, $\rho_0$ and $U_0 = c_v n_0 MT_0$. We assume the local equilibrium at steady state. The sketch of the system and the boundary conditions (BC) is shown in Fig. 1.

Equations 3-5 are solved using COMSOL Multiphysics 5.4. Because the system is symmetric, it is sufficient to consider the planar compressible Poiseuille flow in a half of the channel only. We assume a fully developed inlet velocity profile i.e. the gas has established parabolic velocity profile already at the entrance of the channel - see Fig. 1. The temperature at the inlet is uniform and equals to $T_0$ and it develops along the channel. We also assume no heat flow at the outlet.

Accuracy of the numerical simulations was controlled using two parameters: relative tolerance and the number of mesh elements. The first parameter is a convergence criteria value for iterative solver. Iterative processes within the solver sequence continue to iterate on the solution attempt until the calculated relative error approximation drops below the pre-specified relative tolerance. For the data shown Fig. 2 and Fig. 3 relative tolerance was set to $1 \times 10^{-4}$ and the employed mesh was rectangular with a size of 200 x 500 elements (width x height). Values of internal energy shown in Fig. 4 were extrapolated from three different meshes: 200 x 500, 240 x 600 and 320 x 800 and relative tolerance equal to $1 \times 10^{-8}$.

In the horizontal direction we have distributed the mesh elements in a symmetrical geometric sequence with the symmetry axis at $L_x/2$. For vertical direction, we applied arithmetic distribution. The element size ratio of the largest element in the sequence to the smallest one was set to 10 in both directions. As a result, we obtained the highest refinement at the boundaries of the channel. For the size of the channel $L_x = 0.5$ m and $L_y = 0.05$ m, this yields $\Delta x \approx 4$ mm and $\Delta y \approx 0.1$ mm at the centre, $\Delta x \approx 1.5$ mm at the inlet and the outlet and $\Delta y \approx 0.01$ mm close to the sidewall. The greatest refinement at the boundaries and two types of distributions for both directions allows for more accurate calculation of fluxes and yields higher accuracy of the model in comparison with uniform refinement.

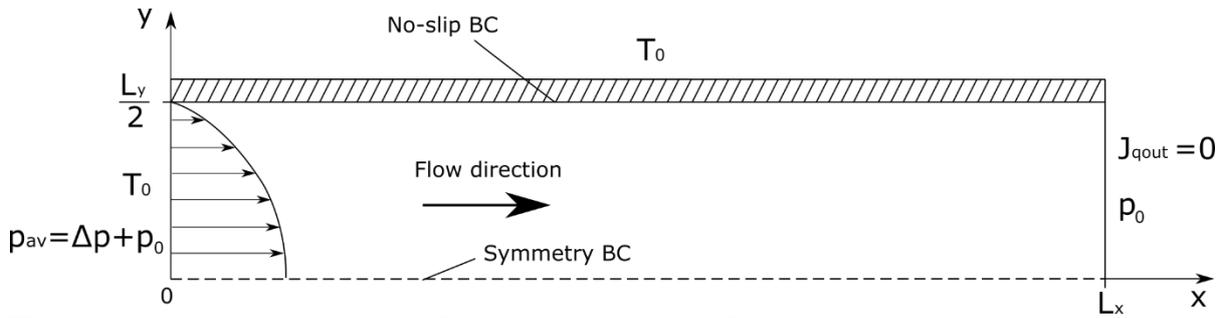

FIG. 1. Geometry and boundary conditions (BC) of the modelled system. Flow profile at the entrance is fully developed with the average pressure set to the value p$_{av}$. $L_y$ and $L_x$ denote width and length of the channel, respectively. Inlet and the channel's sidewalls are thermostated at $T = T_0$. We applied zero heatflux BC across the outlet ($J_{qout} = 0$).

We selected Helium ($^4$He) as a representative of an ideal gas. Characteristic parameters of He, such as $\mu$ and $k$ have been studied experimentally in a broad range of temperatures and pressures. In the main text we present simulation results performed at He parameters for which the Reynolds number ($Re \approx \frac{\Delta p(p_0 + \Delta p)ML_y^3}{12RT\mu^2 L_x}$) is smaller than 2000 to keep the flow in the laminar regime.

A set of parameters characterising a single simulation, and their range applied in this work consists of $c_v = \frac{3}{2}\frac{R}{M}$, $c_p = \frac{5}{2}\frac{R}{M}$, $M = 4.0026$ g mol$^{-1}$, $k$ from 0.05 to 0.65 W m$^{-1}$K$^{-1}$, $\mu$ from $6 \times 10^{-6}$ to $9 \times 10^{-5}$ Pa s, $p_0$ from 0.2 to 2 atm, $\Delta p$ from 0.005 to 50.6 Pa and $T_0$ from 51 to 2500 K. The horizontal and vertical size of the channel varies between 0.3 m $< L_x <$ 8.4 m and 0.024 m $< L_y <$ 0.532 m.

We have assumed that $\mu$ and $k$ for the considered gas depend only on temperature and are increasing functions. Based on this assumption we have interpolated their values from polynomial functions fitted to available experimental data [22,23] (for details see Table I and Table II in Supplemental Material (SM) [24]). We also assumed that $\mu = \mu(T_0)$ and $k = k(T_0)$ since spatial changes of temperature in the flow are small ($T - T_0 < 0.001$ K).

For the considered range of parameters (especially small $\Delta p/p_0$) the medium is weakly compressed. Therefore, similarly to incompressible case, we expect flow profile to be nearly parabolic, yet with small corrections accounting for compressibility. Also due to compressibility small transverse velocity in the system is expected. In case of thermal properties of the flow we expect transition from uniform inlet temperature $T_0$ to a fully developed profile only for the longest channels $L_x \gtrsim 8$m. Note that as a fully developed temperature profile we define scenario in which temperature distribution across the channel does not change in its longitudinal direction $T(x,y) = T(x + dx, y)$.

## III. RESULTS

First, we have validated the model against the analytical solution adopted from [29], where the velocity profile in the flow direction is parabolic:

$$u(y) = \frac{\Delta p}{L_x} \frac{L_y^2}{2\mu} \left[ \frac{1}{4} - \left(\frac{y}{L_y}\right)^2 \right], \tag{9}$$

and the transverse velocity was $v(y) = 0$. The following equation describes the fully developed temperature profile in this case:

$$T(y) = T_0 - \frac{1}{8} \frac{\Delta p^2 L_y^4}{\mu k L_x^2} \left[ \frac{1}{4} - \left(\frac{y}{L_y}\right)^2 \right]^2. \tag{10}$$

In order to obtain a fully developed temperature profile in numerical simulations, we modified inlet BC for temperature i.e. we have imposed periodic boundary conditions for the inlet and outlet temperature. Results for velocity and temperature profiles the fully developed case are shown in Fig. 2(a) and 2(b) respectively. One can see that numerical results agree with the analytical solution for the typical pressure values considered in this paper (error values are given in figure caption). The deviation of flow profiles from the incompressible case is demonstrated in SM.

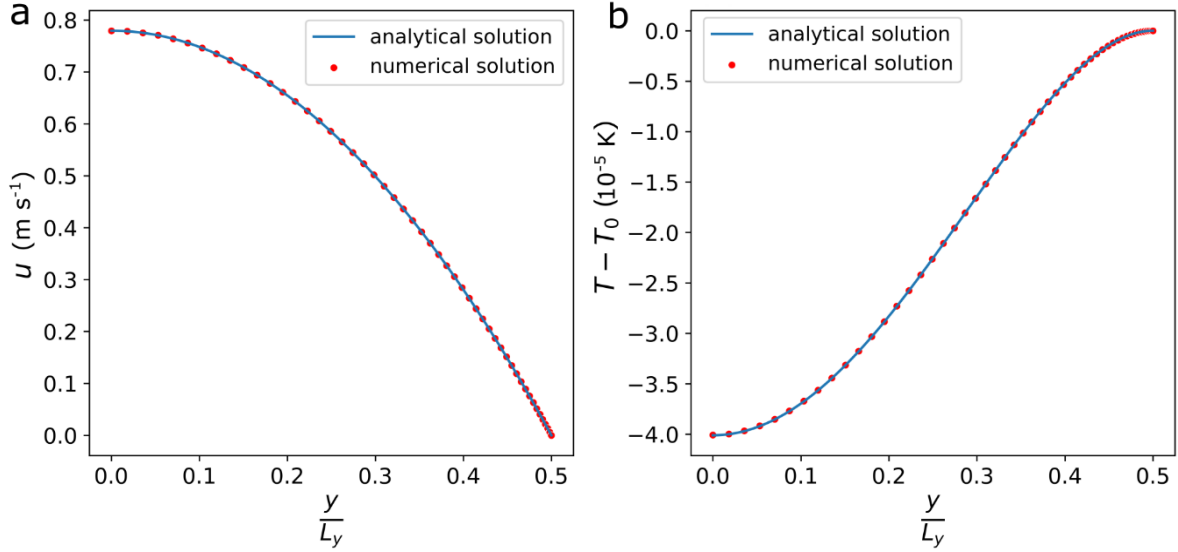

FIG. 2. (a) Velocity in the flow direction and (b) the temperature profile calculated using material parameters for He. The values of other parameters are: $L_x = 0.5$ m, $L_y = 0.025$ m, $T_0 = 300$ K, $\mu = 2.0048 \times 10^{-5}$ Pa s, k = 0.152 W m$^{-1}$ K$^{-1}$, $p_0 = 1$ atm, $\Delta p = 0.5$ Pa. The profiles correspond to a fully developed flow and temperature profiles and are compared to the analytical solution adopted from [29]. The maximal absolute errors equal to (a) $|u|_{err} = 4.5 \times 10^{-5}$ m s$^{-1}$ and (b) $|T - T_0|_{err} = 3.5 \times 10^{-8}$.

**A. Compressible Poiseuille flow of ideal gas – deviation from the analytic internal energy equation**

By applying lubrication approximation and following the methodology presented by Schwartz [30] we obtained analytical solution for the isothermal compressible flow (see SM for details):

$$\frac{U}{U_0} = \frac{1 + \frac{\Delta p}{p_0} + \frac{1}{3}\left(\frac{\Delta p}{p_0}\right)^2}{1 + \frac{1}{2}\frac{\Delta p}{p_0}}. \tag{11}$$

Derivative of internal energy from Eq. (11) w.r.t $\frac{\Delta p}{p_0}$ at small values of $\frac{\Delta p}{p_0}$ is equal to

$$\left.\frac{\partial\left(\frac{U}{U_0}\right)}{\partial\left(\frac{\Delta p}{p_0}\right)}\right|_{\frac{\Delta p}{p_0}=0} = \frac{1}{2}. \tag{12}$$

The above result implies that in the limit of small pressures the ratio $\Delta U/U_0$ of internal energy stored in NESS to the energy of ES depends only on $\frac{\Delta p}{p_0}$ with the proportionality factor 1/2. This result is compared to the dependence of $\Delta U/U_0$ on $\frac{\Delta p}{p_0}$ from the numerical simulation in Fig. 3. Note that each point in the plot corresponds to different simulation parameters. The fitted linear function slightly deviates from the exact $\frac{1}{2}$ slope that indicates that dependence on $\frac{\Delta p}{p_0}$ from Eq. (11) is not sufficient to calculate $U$ precisely. One can also see that for the range of pressures considered in this work, the energy stored in NESS is small (of the order of 0.0001%) with respect to the system's energy in equilibrium. Yet this energy is increasing with $\frac{\Delta p}{p_0}$ and can become significant for large pressure differences $\Delta p$ or low background pressures $p_0$.

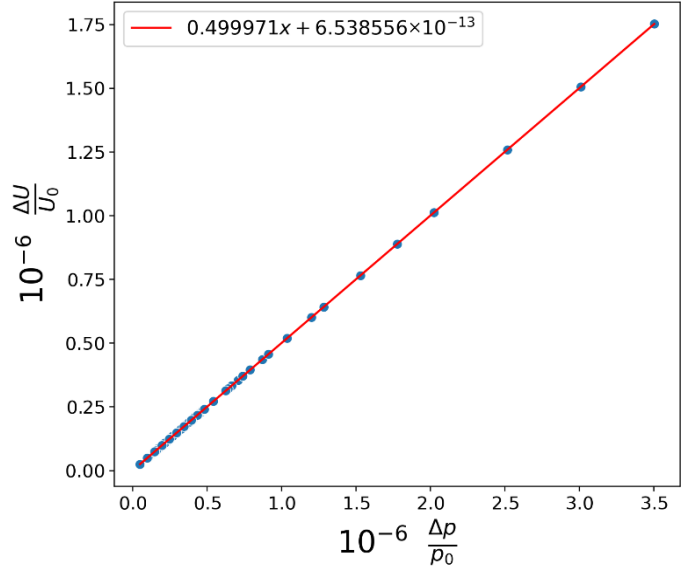

FIG. 3. Dependence of $\Delta U/U_0$ on $\frac{\Delta p}{p_0}$ obtained from numerical model and its linear fit (equation and coefficients shown in the legend). Each point on the plot corresponds to a single simulation with a unique set of parameters varied arbitrarily within the following ranges: $\Delta p$ from 0.001 Pa to 0.08 Pa, $p_0$ from 0.2 atm to 2 atm, $T_0$ from 50 K to 2342 K, $L_x$ from 0.5 m to 8.4 m and $L_y$ from 0.05 m to 0.1 m. The exact algorithm of selecting parameter values is described in SM.

Next, we performed simulations with $\frac{\Delta p}{p_0}$ fixed and all the other parameters changed in the same range as previously in Fig. 3. For this simulation, as described in the previous section, we changed relative tolerance to $1 \times 10^{-8}$ since, at this tolerance, the linear dependency of $U$ on $\frac{\Delta p}{p_0}$ converges to the analytical value with the inversed, squared number of mesh elements. Therefore, the final values of internal energy in Fig. 4 were calculated as the intercept from the linear fit to the data obtained for the three different meshes. The estimation error of $U$, marked with error bars on the plots, is equal to the uncertainty of the intercept.

The results shown in Fig. 3 indicate the existence of variables other than $\frac{\Delta p}{p_0}$, that affect internal energy in the studied system. One can see that despite fixing $\frac{\Delta p}{p_0}$, $U$ still changes with other parameters such as

$T_0$ (Fig. 4(a)) or $p_0$ (Fig. 4(b)). The most substantial correction to the linear law is observed at small temperatures ($T_0$ < 500 K) and channel dimensions (0.5 m < $L_x$ < 1.4 m and 0.05 m < $L_y$ < 0.068 m) as marked with purple colour on the plot (segment 1). Yet these changes are by six orders of magnitude smaller compared to the linear dependency (see scales in Figs. 3 and 4).

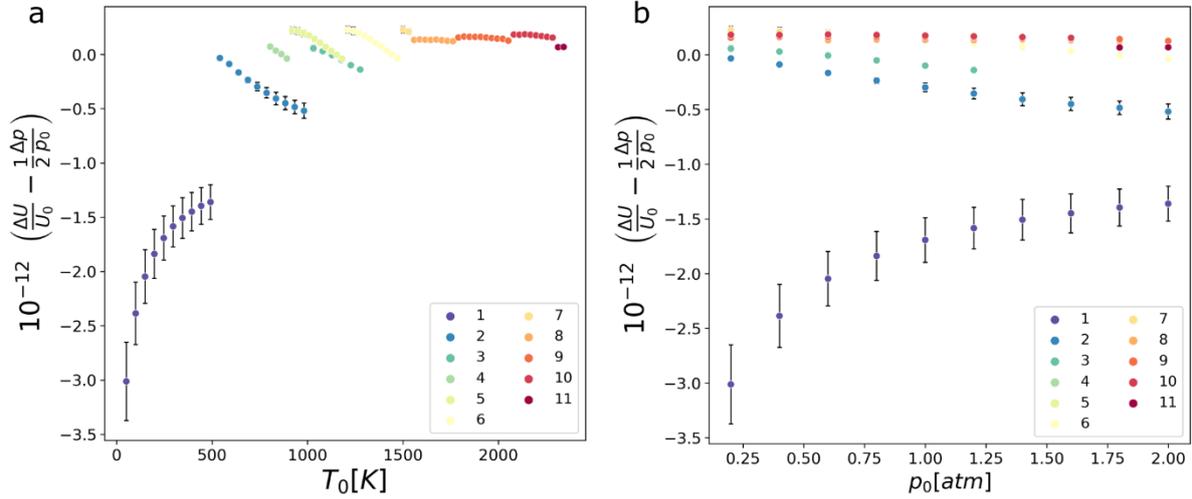

FIG. 4. The results of simulations with $\frac{\Delta p}{p_0}$ fixed at 2.5 × 10$^{-7}$ while the values of all the other parameters were changed in every simulation and varied as follows: $\Delta p$ from 0.001Pa to 0.08 Pa, $p_0$ from 0.2 atm to 2 atm, $T_0$ from 50 K to 2342 K, $L_x$ from 0.5 m to 8.4 m and $L_y$ from 0.05 m to 0.1 m. $\frac{\Delta U}{U_0} - \frac{1}{2}\frac{\Delta p}{p_0}$ is presented as a function of two arbitrarily selected parameters: (a) $T_0$ and (b) $p_0$. Additionally, the separate 11 line segments were colour-coded and enumerated. The range of parameters for each segment is given in Table III in SM.

In order to identify the pressure regime at which additional variables become important, we have performed a set of simulations in which we have varied $\Delta p$ while other simulation parameters have been kept constant. First, let us notice that the slope of $\Delta U/U_0$ as a function of $\frac{\Delta p}{p_0}$ is closer to exact ½ value in this case, as shown in Fig. 5(a). It seems reasonable since all the other parameters, and thus, variables were fixed in this case, and internal energy change should occur only due to linear expression from Eq.

(10). However, as shown in Fig. 5(b) slight discrepancy between analytical and numerical simulation could still be observed.

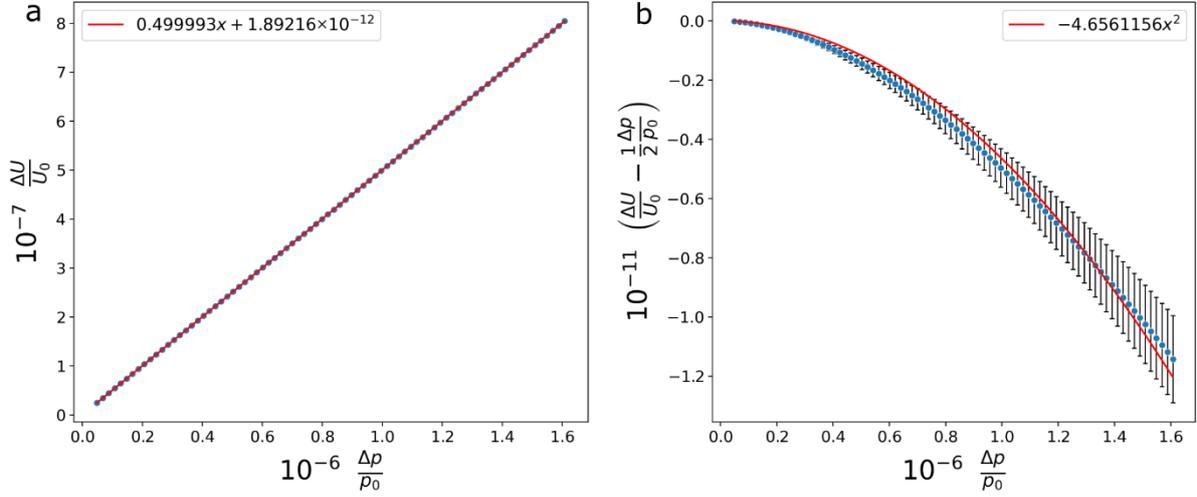

FIG. 5. (a) $\frac{\Delta U}{U_0}$ and (b) $\frac{\Delta U}{U_0} - \frac{1}{2}\frac{\Delta p}{p_0}$ in function of $\Delta p/p_0$. $\Delta p$ was changed from 0.005 Pa to 0.163 Pa. The values of other simulation parameters are constant and equal to $p_0$=1 atm, $T_0$=600 K, $L_x$=0.5 m, $L_y$=0.05m. The red line in is the best quadratic fit with the equation and coefficients shown in the legend.

One can see that there is an additional, negative contribution to the stored energy that is a quadratic function of $\frac{\Delta p}{p_0}$. Note that this contribution is much smaller than the total stored energy (of the order of 0.0015% compared to the linear term). Yet, as it increases quadratically with $\frac{\Delta p}{p_0}$ it might become essential for systems exposed to large pressure differences or at close to vacuum conditions. This result also suggests that for compressible flow Eq. (11) should have the following form:

$$U = U_0\left[1 + \frac{1}{2}\frac{\Delta p}{p_0} - \left(\frac{\Delta p}{p_0}\right)^2 g\right], \tag{13}$$

where $g$ is a function of additional variables that have not yet been identified. It is worth noting that in Eq. (11), derived based on lubrication approximation, $g$ is constant ($g = 1/12$) so it does not include contribution from effects associated with eg. temperature profile.

**B. Natural variables of the internal energy function**

In order to find the variables of function $g$, we rewrite Eqs. (3-5) by normalising the variables with characteristic quantities describing the system in equilibrium:

$$u^* = \frac{u}{\sqrt{\frac{RT_0}{M}}}, \rho^* = \frac{\rho}{\rho_0}, p^* = \frac{p}{p_0}, x^* = \frac{x}{L_x}, y^* = \frac{y}{L_y}.$$

As demonstrated in Fig. S1, the transverse velocity components in the considered flow for applied pressures are small. Thus for simplicity, we assume that the velocity field has only one component $\vec{u} = (u(x,y), 0, 0)$. Note that this assumption is only used in derivation of scaled equations and not for numerical simulations. Finally, the dimensionless equations have the following form (for detailed scaling procedure see SM):

$$\frac{L_y}{L_x}\frac{\partial(\rho^* u^*)}{\partial x^*} = 0 \tag{14}$$

$$\frac{p_0 L_x}{\mu}\sqrt{\frac{M}{RT_0}} \rho^* u^* \frac{\partial u^*}{\partial x^*} \tag{15}$$

$$= -\frac{p_0 L_x}{\mu}\sqrt{\frac{M}{RT_0}}\frac{\partial p^*}{\partial x^*} + \frac{\partial^2 u^*}{\partial (x^*)^2} + \frac{L_x^2}{L_y^2}\frac{\partial^2 u^*}{\partial (y^*)^2} + \frac{1}{3}\frac{\partial^2 u^*}{\partial (x^*)^2}$$

$$+ \frac{1}{3}\frac{L_x}{L_y}\frac{\partial^2 u^*}{\partial x^* \partial y^*}$$

$$\frac{p_0 L_x}{\mu}\sqrt{\frac{M}{RT_0}} \rho^* u^* \frac{\partial T^*}{\partial x^*} \tag{16}$$

$$= \frac{k}{\mu c_p}\frac{\partial^2 T^*}{\partial (x^*)^2} + \frac{k}{\mu c_p}\frac{L_x^2}{L_y^2}\frac{\partial^2 T^*}{\partial (y^*)^2} + \frac{p_0 L_x}{\mu}\sqrt{\frac{M}{RT_0}}\frac{2}{5}u^*\frac{\partial p^*}{\partial x^*}$$

$$+ \frac{4}{5}\left(\frac{\partial u^*}{\partial x^*}\right)^2 + \frac{2}{5}\left(\frac{\partial u^*}{\partial y^*}\right)^2 - \frac{4}{15}\left(\frac{\partial u^*}{\partial x^*}\right)^2.$$

In these equations three dimensionless numbers appear: $\frac{p_0 L_x}{\mu}\sqrt{\frac{M}{RT_0}}, \frac{L_y}{L_x}$ and $\frac{\mu c_p}{k} = Pr$. The first one is the analogue of the Reynolds number. The second one describes anisotropy of the system, and the third one is the Prandtl number ($Pr$). One can demonstrate, that combination of these numbers and the primary variable $\frac{\Delta p}{p_0}$ yields the variable for $U$ of the incompressible system from Eq. (2):

$$\frac{\Delta p^2 L_y^4}{\mu k T_0 L_x^2} = \left(\frac{\Delta p}{p_0}\right)^2 \cdot \left(\frac{p_0 L_x}{\mu}\sqrt{\frac{M}{RT_0}}\right)^2 \cdot \left(\frac{L_y}{L_x}\right)^4 \cdot \frac{2}{5} Pr. \qquad (17)$$

Furthermore, the quadratic dependence on $\frac{\Delta p}{p_0}$ agrees with the result for compressible fluid shown in Fig. 5(b). As the result, we expect that the Eq. (13) has four independent variables and in a general form can be expressed as follows:

$$U = U_0 \left[1 + \frac{1}{2}\frac{\Delta p}{p_0} + \left(\frac{\Delta p}{p_0}\right)^2 g\left(\frac{p_0 L_x}{\mu}\sqrt{\frac{M}{RT_0}}, \frac{L_y}{L_x}, Pr\right)\right]. \qquad (18)$$

In order to verify if the proposed function variables from Eq. (18) are correct, we performed four series of simulations – each with one of the four variables changed, and the remaining ones kept constant. Properly selected variable, when changed, gives a smooth curve for all the other variables kept constant. Improper selection of variable leads to the formation of branches on the internal energy graph. Properly and improperly selected variables are shown in Fig. 6(a) and (b) respectively.

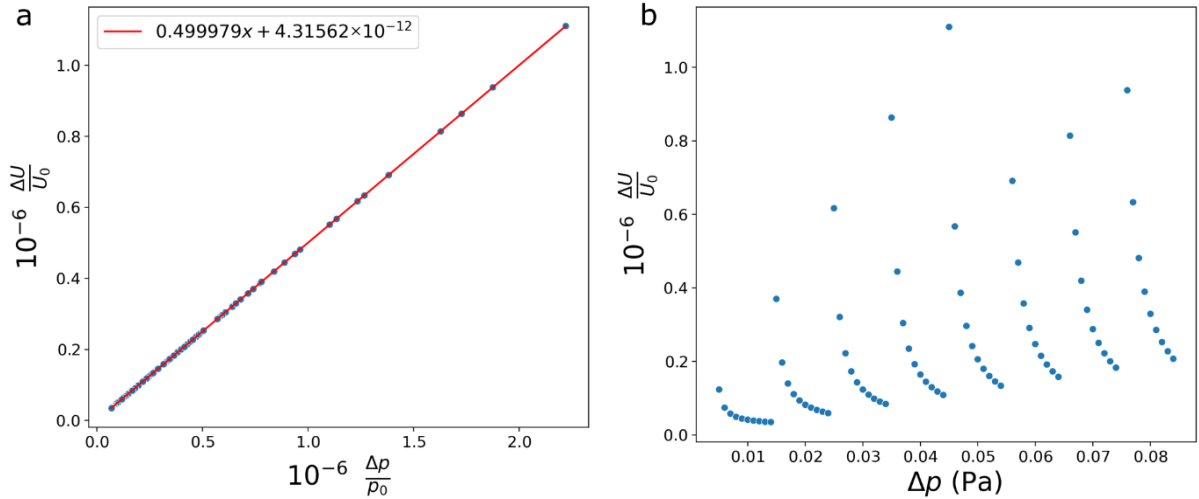

FIG. 6. $\Delta U/U_0$ in the function of (a) $\Delta p/p_0$ and (b) $\Delta p$. The values of other dimensionless parameters are constant and equal to $\frac{p_0 L_x}{\mu}\sqrt{\frac{M}{RT_0}} = 3 \times 10^6$, $\frac{L_y}{L_x} = 0.1$, $Pr = 0.67$. The red line in (a) is the best linear fit with the equation and coefficients shown in the legend. Note that $\Delta p$ is not a proper variable for $\Delta U/U_0$.

We have applied the same procedure to the other variables from Eq. (18), namely: $\frac{L_y}{L_x}$, $\frac{p_0 L_x}{\mu}\sqrt{\frac{M}{RT_0}}$ and $Pr$. This time however we were only interested in the contribution to $U$ from the term dependent on additional variables. Thus we have subtracted the constant value $\frac{1}{2}\frac{\Delta p}{p_0}$ from the ratio of stored to ES internal energy and traced the following quantity: $\frac{\Delta U}{U_0} - \frac{1}{2}\frac{\Delta p}{p_0}$. We present the respective charts in Fig. 7(a), Fig. 8(a) and Fig. 9(a) along with the fitting curves for $\frac{L_y}{L_x}$ and $\frac{p_0 L_x}{\mu}\sqrt{\frac{M}{RT_0}}$ dependence. For comparison, in all the cases dependence on an arbitrarily chosen variable is shown in panel (b).

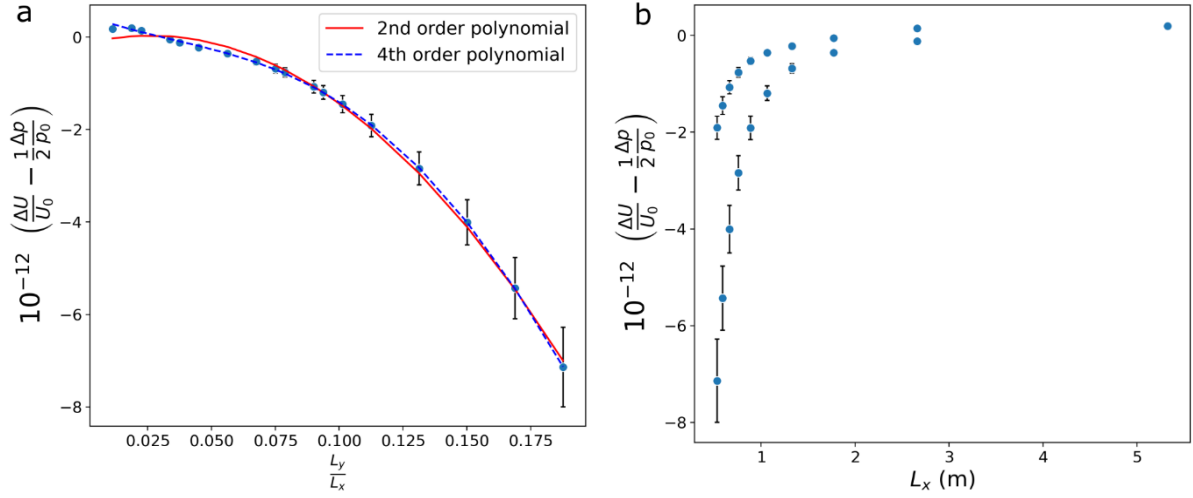

FIG. 7. $\frac{\Delta U}{U_0} - \frac{1}{2}\frac{\Delta p}{p_0}$ in the function of (a) $\frac{L_y}{L_x}$ and (b) $L_x$. The values of other dimensionless parameters are constant and equal to $\frac{\Delta p}{p_0} = 2.5 \times 10^{-7}$, $\frac{p_0 L_x}{\mu}\sqrt{\frac{M}{RT_0}} = 3 \times 10^6$, $Pr = 0.67$. The solid red and dashed blue lines in (a) are the best fit of a second and a fourth-degree polynomial function. The exact equations and polynomial coefficients are given in Table IV in SM. Note that $L_x$ is not a proper variable for $\Delta U/U_0$.

In agreement with Eq. (17), the dependence of $U$ on $L_y/L_x$ can be described with a fourth-order polynomial. Fitting with quadratic function also gives a good agreement; however, a slight discrepancy is visible at small channel aspect ratio. The dependence on $\frac{p_0 L_x}{\mu}\sqrt{\frac{M}{RT_0}}$ is also more accurately described by the fourth than by the second-order polynomial as shown in Fig. 8(a). In this case, however, based on incompressible flow solution, we have expected that quadratic function would be used. Parameters of the fitted polynomials from Fig. 7 and Fig. 8 are given in SM.

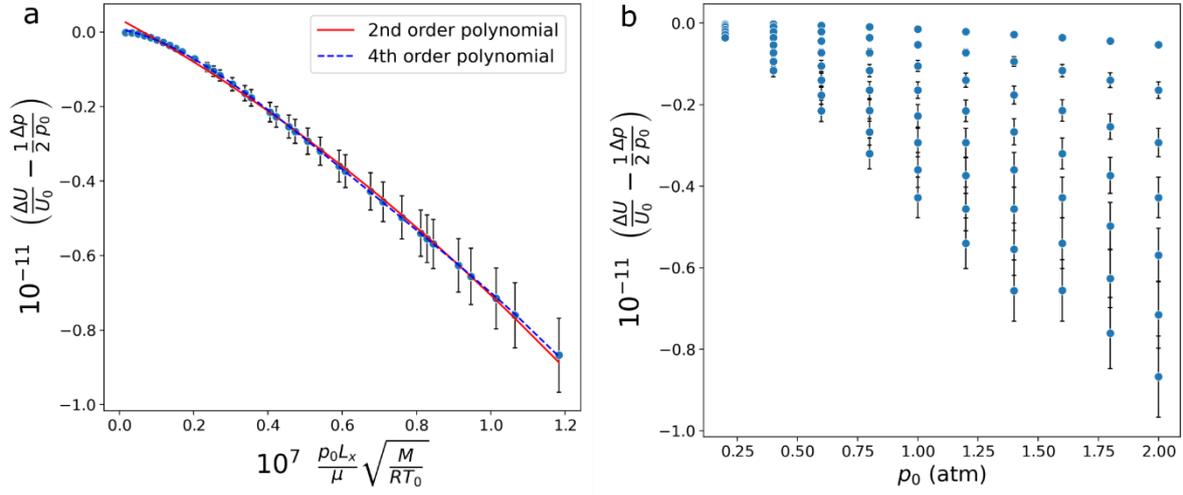

FIG. 8. $\frac{\Delta U}{U_0} - \frac{1}{2}\frac{\Delta p}{p_0}$ in the function of (a) $\frac{p_0 L_x}{\mu}\sqrt{\frac{M}{RT_0}}$ and (b) $p_0$. The values of other dimensionless parameters are constant and equal to $\frac{\Delta p}{p_0} = 2.47 \times 10^{-7}, \frac{L_y}{L_x} = 0.1, Pr = 0.67$. The solid red and dashed blue lines in (a) are the best fit of a second and a fourth-degree polynomial function. The exact equations and polynomial coefficients are given in Table IV in SM. Note that $p_0$ is not a proper variable for $\Delta U/U_0$.

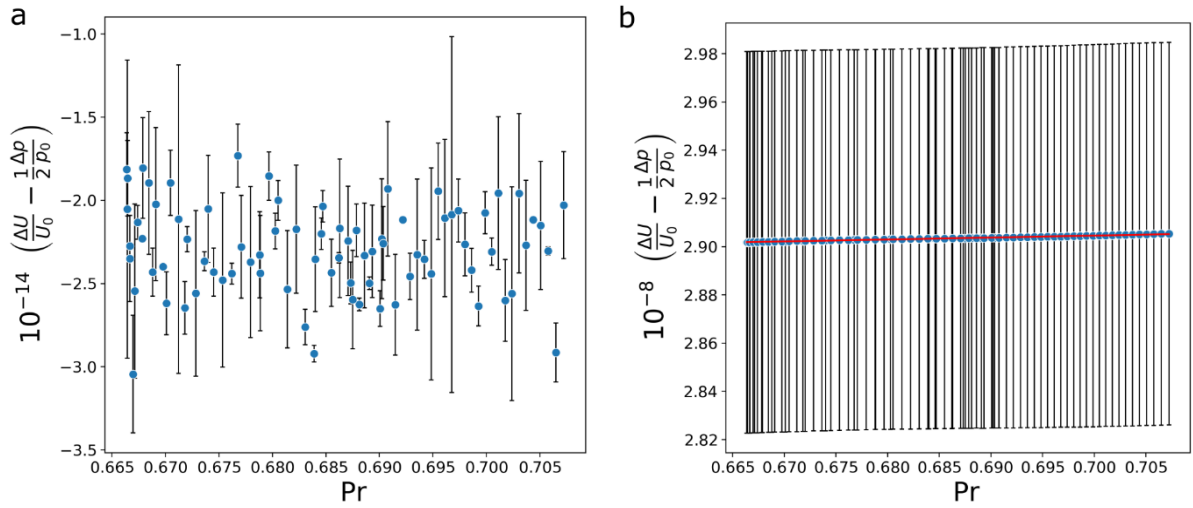

FIG. 9. $\frac{\Delta U}{U_0} - \frac{1}{2}\frac{\Delta p}{p_0}$ in the function of $Pr$ for $\frac{\Delta p}{p_0}$ fixed at (a) $2.5 \times 10^{-7}$ and (b) $2.5 \times 10^{-4}$. The values of other dimensionless parameters are constant the same for both cases: $\frac{p_0 L_x}{\mu}\sqrt{\frac{M}{RT_0}} = 3 \times 10^6, \frac{L_y}{L_x} = 0.1$.

As shown in Fig. 9(a) we have observed that $\frac{\Delta U}{U_0} - \frac{1}{2}\frac{\Delta p}{p_0}$ is independent of $Pr$ number for the range of parameters used previously (Figs. 6-8). Therefore for moderate values of $\frac{\Delta p}{p_0}$ ($< 10^{-4}$), Eq. (18) can be simplified to the following form:

$$U = U_0\left[1 + \frac{1}{2}\frac{\Delta p}{p_0} + \left(\frac{\Delta p}{p_0}\right)^2 g\left(\frac{p_0 L_x}{\mu}\sqrt{\frac{M}{RT_0}}, \frac{L_y}{L_x}\right)\right]. \quad (19)$$

## IV. CONCLUSIONS

In this paper, we have shown that generally the internal energy of compressible Poiseuille flow of ideal gas in a planar geometry has the following structure:

$$U = U_0 f\left(\frac{\Delta p}{p_0}, \frac{p_0 L_x}{\mu}\sqrt{\frac{M}{RT_0}}, \frac{L_y}{L_x}, Pr\right). \quad (20)$$

However for pressure drop between inlet and outlet of the channel and absolute pressure between 0.2 and 2 atm considered in this work, dependence on $Pr$ can be neglected, and the function from Eq. (20) can be approximated as $f = \frac{1}{2}\frac{\Delta p}{p_0} + \left(\frac{\Delta p}{p_0}\right)^2 g\left(\frac{p_0 L_x}{\mu}\sqrt{\frac{M}{RT_0}}, \frac{L_y}{L_x}\right)$. The first and dominant argument of the internal energy function is a linear term $\frac{\Delta p}{p_0}$ and it is associated with the compressibility of the medium. In this case, externally applied pressure results in pressure gradient along the tube but also, according to the equation of state, increases the density of the medium and as a consequence, its internal energy. This becomes evident in comparison with the incompressible scenario where this term vanishes and only quadratic dependence on $\frac{\Delta p}{p_0}$ is present. We have demonstrated that the contribution to the energy stored in NESS attributed to $\frac{1}{2}\frac{\Delta p}{p_0}$ is six orders of magnitude larger compared to the changes associated with the other variables. Furthermore, the quadratic term vanishes as $g$ converges to 0 i.e. for small values of $\frac{L_y}{L_x}$ and $\frac{p_0 L_x}{\mu}\sqrt{\frac{M}{RT_0}}$ as for this regime velocity of the flow is small (see Fig. S2 in SM) and only static compression of the gas contributes to internal energy change. Thus, unless for large $\Delta p$ or close to vacuum conditions, the energy stored in the compressible Poiseuille flow of ideal gas can be calculated with good approximation as $\Delta U = \frac{1}{2}\frac{\Delta p}{p_0}$. This seems especially important if the hypothesis of a connection between $\Delta U$ and macroscopic properties of the system is to be explored experimentally.

## ACKNOWLEDGMENTS

This research was founded by NCN within Sonata grant 2019/35/D/ST5/03613.

**Supplemental Material for "Internal energy in compressible Poiseuille flow"**

Konrad Gizynski[1], Karol Makuch[1], Jan Paczesny[1], Yirui Zhang[1], Anna Maciołek[1,2] and Robert Holyst[1*]

[1]Institute of Physical Chemistry, Polish Academy of Sciences
Kasprzaka 44/52, 01-224 Warsaw, Poland

[2]Max-Planck-Institut für Intelligente Systeme,
Heisenbergstr. 3, D-70569 Stuttgart, Germany

TABLE I. Coefficients of polynomial function: $\mu(T) = a_1 + a_2 T + a_3 T^2 + a_4 T^3 + a_5 T^4$ describing viscosity of Helium gas for two temperature ranges.

| Temperature range | $a1$ | $a2$ | $a3$ | $a4$ | $a5$ |
|---|---|---|---|---|---|
| $50 < T <= 313$ | $1.482 \times 10^{-6}$ | $1.037 \times 10^{-7}$ | $-2.798 \times 10^{-10}$ | $6.514 \times 10^{-13}$ | $-6.105 \times 10^{-16}$ |
| $313 < T <= 2500$ | $6.070 \times 10^{-6}$ | $5.039 \times 10^{-8}$ | $-1.341 \times 10^{-11}$ | $3.701 \times 10^{-15}$ | $-4.109 \times 10^{-19}$ |

TABLE II. Coefficients of polynomial function: $k(T) = b_1 + b_2 T + b_3 T^2 + b_4 T^3 + b_5 T^4$ describing thermal conductivity of Helium gas for two temperature ranges.

| Temperature range | $b1$ | $b2$ | $b3$ | $b4$ | $b5$ |
|---|---|---|---|---|---|
| $50 < T <= 280$ | $1.090 \times 10^{-2}$ | $8.302 \times 10^{-4}$ | $-2.877 \times 10^{-6}$ | $8.838 \times 10^{-9}$ | $-1.093 \times 10^{-11}$ |
| $280 < T <= 2500$ | $2.707 \times 10^{-2}$ | $4.686 \times 10^{-4}$ | $-1.945 \times 10^{-7}$ | $6.493 \times 10^{-11}$ | $-8.723 \times 10^{-15}$ |

**The algorithm for selecting parameter values used for data shown in Fig. 3:**

$p_0^n = 0.2 \text{ atm} + (n \% 10) * 0.2 \text{ atm}$ (for n=0,1….79)

$\Delta p^n = 0.001 \text{ Pa} + n * 0.001 \text{ Pa}$ (for n=0,1….79)

$T_0^n = 51 K + n * 49 \text{ K}$ (for n=0,1….25) and $T_0^n = 805 \text{ K} + (n - 26) * 29 \text{ K}$ (for n=26,27….79)

$L_x^n = 0.5 \text{ m} + n * 0.1 \text{ m}$ (for n=0,1,2....79)

$L_y^n = 0.05 \text{ m} + (n \% 26) * 0.002 \text{ m}$ (for n=0,1,2....79)

Operation % (modulo) returns division remainder of the two operands eg. 4 % 3 = 1.

The values of µ and k were calculated based on functions from Table I and Table II.

TABLE III. Range of parameters for line segments (1-11) shown in Fig. 4. The points within a segment are distributed uniformly with the following spacing: $\Delta p_0$ = 0.2 atm, $\Delta(\Delta p)$ = 0.005066 Pa, $\Delta T_0$ = 49 K, $\Delta L_x$ = 0.1 m, $\Delta L_y$ = 0.002 m. Viscosity and thermal conductivity were interpolated based on $T_0$ using polynomial functions shown in Table I and Table II respectively.

| Segment no. | $p_0$ (atm) | $\Delta p$ (Pa) | $T_0$ (K) | $L_x$ (m) | $L_y$ (m) |
|---|---|---|---|---|---|
| 1 | 0.2 – 2 | 0.00506625 - 0.0506625 | 51 - 492 | 0.5-1.4 | 0.05-0.68 |
| 2 | 0.2 – 2 | 0.00506625 - 0.0506625 | 541 - 982 | 1.5-2.4 | 0.07-0.088 |
| 3 | 0.2 – 1.2 | 0.00506625 - 0.0303975 | 1031 -1276 | 2.5-3 | 0.09-0.1 |
| 4 | 1.4 – 2 | 0.03546375 - 0.0506625 | 805 – 892 | 3.1-3.4 | 0.05-0.056 |
| 5 | 0.2 – 2 | 0.00506625 - 0.0506625 | 921 – 1182 | 3.5-4.4 | 0.058-0.076 |
| 6 | 0.2 – 2 | 0.00506625 - 0.0506625 | 1211 – 1472 | 4.5-5.4 | 0.078 – 0.096 |
| 7 | 0.2 – 0.4 | 0.00506625 - 0.0101325 | 1501 – 1530 | 5.5-5.6 | 0.098-0.1 |
| 8 | 0.6 – 2 | 0.01519875- 0.0506625 | 1559-1762 | 5.7-6.4 | 0.05-0.064 |
| 9 | 0.2 – 2 | 0.00506625 - 0.0506625 | 1791 – 2052 | 6.5-7.4 | 0.066 – 0.084 |
| 10 | 0.2 – 1.6 | 0.00506625 - 0.04053 | 2081-2284 | 7.5-8.2 | 0.086 – 0.1 |
| 11 | 1.8 – 2 | 0.04559625 - 0.0506625 | 2313 - 2342 | 8.3-8.4 | 0.5-0.52 |

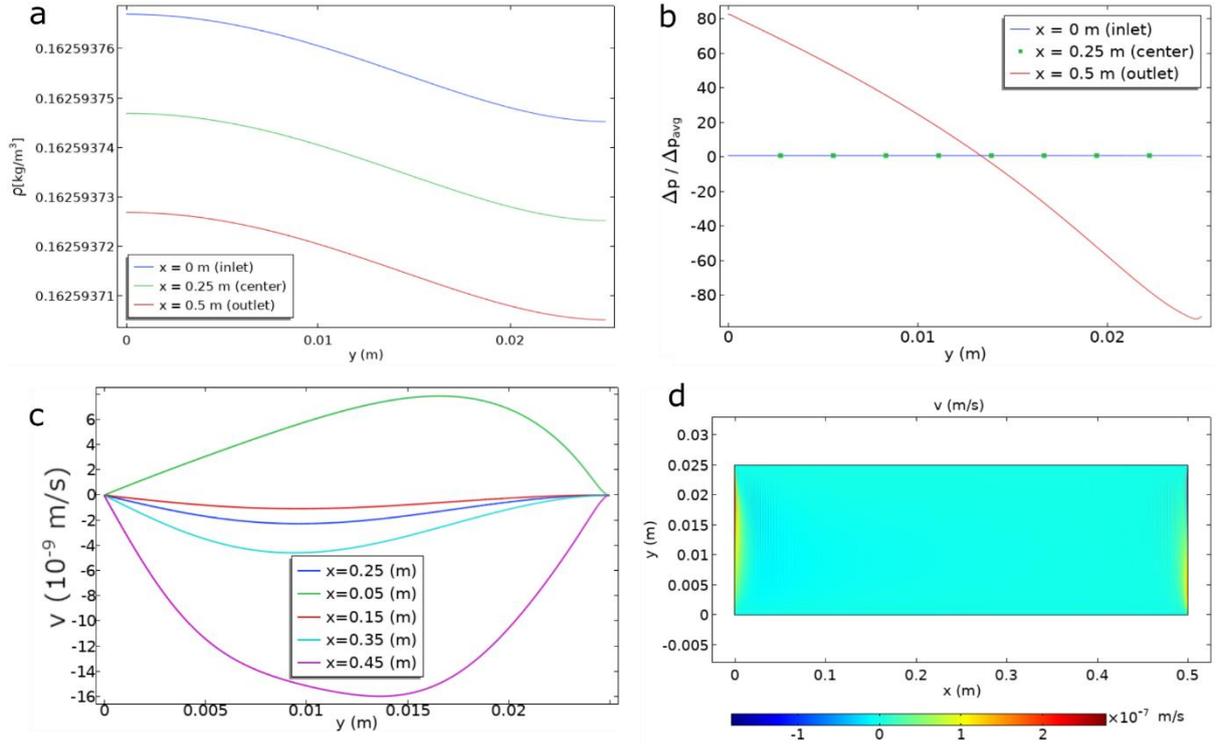

FIG. S1. Profiles of (a) density, (b) pressure and (c) transverse velocity at different positions along the channel. (d) Spatial pressure distribution in the channel.

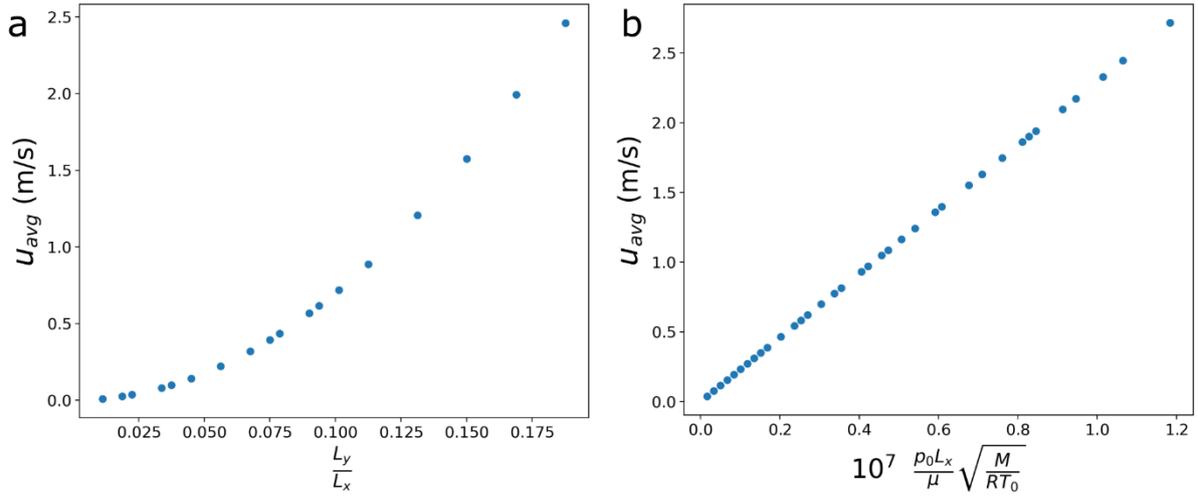

FIG. S2. Average velocity $u_{avg}$ of the gas in function of (a) $\frac{L_y}{L_x}$ and (b) $\frac{p_0 L_x}{\mu}\sqrt{\frac{M}{RT_0}}$.

TABLE IV. Coefficients of polynomial functions $c_1 x^4 + c_2 x^3 + c_3 x^2 + c_4 x + c_5$ used to fit data in Fig. 7 and Fig. 8.

|  | Polynomial | c1 | c2 | c3 | c4 | c5 |
|---|---|---|---|---|---|---|
| Figure 7 | 2-nd order | 0 | 0 | $-2.70 \cdot 10^{-10}$ | $1.42 \cdot 10^{-11}$ | $-1.57 \cdot 10^{-13}$ |
|  | 4-th order | $4.92 \cdot 10^{-9}$ | $-2.89 \cdot 10^{-9}$ | $2.70 \cdot 10^{-10}$ | $-2.21 \cdot 10^{-11}$ | $5 \cdot 10^{-13}$ |
| Figure 8 | 2-nd order | 0 | 0 | $-2.07 \cdot 10^{-26}$ | $-5.34 \cdot 10^{-19}$ | $3.48 \cdot 10^{-13}$ |
|  | 4-th order | $-4.54 \cdot 10^{-40}$ | $1.28 \cdot 10^{-32}$ | $-1.36 \cdot 10^{-25}$ | $-1.73 \cdot 10^{-19}$ | $7.6 \cdot 10^{-14}$ |

**Derivation of analytical expression for internal energy in compressible flow**

We assume parabolic velocity profile for every $x$ position along the tube with an unknown factor $u_{av}$ dependent on $x$:

$$u(x,y) = u_{av}(x) \frac{\left[\left(\frac{L_y}{2}\right)^2 - y^2\right]}{\left(\frac{L_y}{2}\right)^2} = u_{av}(x)\left(1 - \frac{4y^2}{L_y^2}\right). \tag{S1}$$

For zero-th order solution we can neglect dependence of density on $y$: $\rho(x,y) = \rho(x)$ and the constant mass flow rate in each half-channel can be expressed as:

$$\dot{m} = \int_0^{\frac{L_y}{2}} \rho(x) u(x,y) dy = \rho(x) u_{av}(x) \left[y - \frac{4y^3}{3L_y^2}\right]_0^{\frac{L_y}{2}} = \rho(x) u_{av}(x) \frac{L_y}{3}. \tag{S2}$$

For the flows changing slowly in the flow direction nonlinear terms in Navier-Stokes equation can be neglected:

$$u\frac{\partial u}{\partial x} = 0$$

$$v\frac{\partial u}{\partial y} = 0$$

And then x-component of the equation is reduced to:

$$\frac{\partial p}{\partial x} = \mu\left(\frac{\partial^2 u}{\partial x^2} + \frac{\partial^2 u}{\partial y^2}\right) + \frac{1}{3}\mu\frac{\partial^2 u}{\partial x^2}$$

$$\frac{\partial p}{\partial x} = \mu\left(\frac{4}{3}\frac{\partial^2 u}{\partial x^2} + \frac{\partial^2 u}{\partial y^2}\right)$$

$$\frac{\partial p}{\partial x} = \mu\left[\frac{4}{3}\left(1 - \frac{4y^2}{L_y^2}\right)\frac{\partial^2 u_{av}(x)}{\partial x^2} + \frac{8 u_{av}(x)}{L_y^2}\right]$$

By expressing pressure from with the equation of state:

$$p = \frac{RT}{M}\rho(x)$$

one can obtain the following differential equation

$$\frac{RT}{M}\frac{\partial \rho(x)}{\partial x} = \mu\left[\frac{4}{L_y}\left(1 - \frac{4y^2}{L_y^2}\right)\frac{\partial^2 \left(\frac{\dot{m}}{\rho(x)}\right)}{\partial x^2} + \frac{24\dot{m}}{\rho(x)L_y^3}\right]. \tag{S3}$$

Which at the wall $\left(y = \frac{L_y}{2}\right)$ can be simplified to the following form:

$$\frac{d\rho(x)}{dx} = \frac{\mu M}{RT}\frac{24\dot{m}}{\rho(x)L_y^3}. \tag{S4}$$

The final equation for density in function of channel length:

$$\rho(x) = \frac{M(p_0 + \Delta p)}{RT}\sqrt{\left(-\frac{\Delta p(2p_0 + \Delta p)}{L_x(p_0 + \Delta p)^2}\right)x + 1} \tag{S5}$$

From the equation above and the second equation of ideal gas one can derive expression for internal energy in compressible flow:

$$\begin{aligned}
U &\quad \text{(S6)}\\
&= \int_0^{L_x}\int_{-\frac{L_y}{2}}^{\frac{L_y}{2}} c_v \rho(x) T \, dy\, dx\\
&= \int_0^{L_x}\int_{-\frac{L_y}{2}}^{\frac{L_y}{2}} \frac{3}{2}(p_0+\Delta p)\sqrt{\left(-\frac{\Delta p(2p_0+\Delta p)}{L_x(p_0+\Delta p)^2}\right)x+1}\, dy\, dx\\
&= \frac{3}{2}(p_0+\Delta p)L_y \int_0^{L_x}\sqrt{\left(-\frac{\Delta p(2p_0+\Delta p)}{L_x(p_0+\Delta p)^2}\right)x+1}\, dx\\
&= \frac{3}{2}(p_0+\Delta p)L_y \left[\frac{2\left[\left(-\frac{\Delta p(2p_0+\Delta p)}{L_x(p_0+\Delta p)^2}\right)x+1\right]^{\frac{3}{2}}}{3\left[-\frac{\Delta p(2p_0+\Delta p)}{L_x(p_0+\Delta p)^2}\right]}\right]_0^{L_x} =\\
&= \frac{3}{2}(p_0+\Delta p)L_y \left[\frac{2\left\{\left[-\frac{\Delta p(2p_0+\Delta p)}{L_x(p_0+\Delta p)^2}\right]L_x+1\right\}^{\frac{3}{2}}}{3\left[-\frac{\Delta p(2p_0+\Delta p)}{L_x(p_0+\Delta p)^2}\right]} - \frac{2}{3\left[-\frac{\Delta p(2p_0+\Delta p)}{L_x(p_0+\Delta p)^2}\right]}\right]\\
&= -\frac{L_y L_x (p_0+\Delta p)^3}{\Delta p(2p_0+\Delta p)}\left\{\left[\left(-\frac{\Delta p(2p_0+\Delta p)}{(p_0+\Delta p)^2}\right)+1\right]^{\frac{3}{2}}-1\right\}\\
&= -\frac{2U_0(p_0+\Delta p)^3}{3\Delta p\, p_0(2p_0+\Delta p)}\left\{\left[\left(-\frac{\Delta p(2p_0+\Delta p)}{(p_0+\Delta p)^2}\right)+1\right]^{\frac{3}{2}}-1\right\}\\
&= -\frac{2U_0(p_0+\Delta p)^3}{3\Delta p\, p_0(2p_0+\Delta p)}\left\{\left(\frac{-2p_0\Delta p - \Delta p^2 + p_0^2 + 2p_0\Delta p + \Delta p^2}{(p_0+\Delta p)^2}\right)^{\frac{3}{2}}-1\right\}\\
&= \frac{2U_0}{3\Delta p\, p_0(2p_0+\Delta p)}(3p_0^2\Delta p + 3\Delta p^2 p_0 + \Delta p^3)\\
&= \frac{2U_0\left(3+\frac{\Delta p}{p_0}\left(3+\frac{\Delta p}{p_0}\right)\right)}{3\left(2+\frac{\Delta p}{p_0}\right)} = \frac{U_0\left(1+\frac{\Delta p}{p_0}+\frac{1}{3}\left(\frac{\Delta p}{p_0}\right)^2\right)}{1+\frac{1}{2}\frac{\Delta p}{p_0}}.
\end{aligned}$$

**Equations scaling**

1. Continuity equation for 2D compressible flow

$$\nabla(\rho \vec{u}) = 0, \quad \text{(S7)}$$

$$\frac{\partial(\rho u)}{\partial x} + \frac{\partial(\rho v)}{\partial y} = 0. \quad \text{(S8)}$$

After scaling by $\rho_0, p_0, \sqrt{\frac{RT_0}{M}}, L_x, L_y$ such that

$$u^* = \frac{u}{\sqrt{\frac{RT_0}{M}}},$$

$$v^* = \frac{v}{\sqrt{\frac{RT_0}{M}}}$$

$$\rho^* = \frac{\rho}{\rho_0}$$

$$p^* = \frac{p}{p_0}$$

$$x^* = \frac{x}{L_x}$$

$$y^* = \frac{y}{L_y}$$

we get

$$\rho_0 \sqrt{\frac{RT_0}{M}} \left( \frac{1}{L_x} \frac{\partial (\rho^* u^*)}{\partial x^*} + \frac{1}{L_y} \frac{\partial (\rho^* v^*)}{\partial y^*} \right) = 0. \tag{S9}$$

Dividing both sides by $\frac{\rho_0}{L_y}\sqrt{\frac{RT_0}{M}}$ yields

$$\frac{L_y}{L_x} \frac{\partial (\rho^* u^*)}{\partial x^*} + \frac{\partial (\rho^* v^*)}{\partial y^*} = 0. \tag{S10}$$

For Poiseuille flow $v = 0$ and the equation reduces to

$$\frac{L_y}{L_x} \frac{\partial (\rho^* u^*)}{\partial x^*} = 0. \tag{S11}$$

2. Navier-Stokes stationary momentum equation for 2D compressible flow

$$\rho \vec{\mathbf{u}} \cdot \nabla \vec{\mathbf{u}} = -\nabla p + \nabla \cdot \left\{ \mu \left[ \nabla \vec{\mathbf{u}} + (\nabla \vec{\mathbf{u}})^T - \frac{2}{3} (\nabla \cdot \vec{\mathbf{u}}) \mathbf{I} \right] + \vartheta (\nabla \cdot \vec{\mathbf{u}}) \mathbf{I} \right\} \tag{S12}$$

where $\vartheta$ is the second viscosity coefficient (bulk viscosity).

For monoatomic ideal gas $\vartheta = 0$ (Refs. [1–4]). Then

$$\rho \vec{\mathbf{u}} \cdot \nabla \vec{\mathbf{u}} = -\nabla p + \nabla \cdot \left\{ \mu \left[ \nabla \vec{\mathbf{u}} + (\nabla \vec{\mathbf{u}})^T - \frac{2}{3} (\nabla \cdot \vec{\mathbf{u}}) \mathbf{I} \right] \right\}. \tag{S13}$$

For 2d flow in Cartesian coordinates:

$$\rho \left( u \frac{\partial u}{\partial x} + v \frac{\partial u}{\partial y} \right) = -\frac{\partial p}{\partial x} + \mu \left( \frac{\partial^2 u}{\partial x^2} + \frac{\partial^2 u}{\partial y^2} \right) + \frac{1}{3} \mu \frac{\partial}{\partial x} \left( \frac{\partial u}{\partial x} + \frac{\partial v}{\partial y} \right), \tag{S14}$$

$$\rho \left( u \frac{\partial v}{\partial x} + v \frac{\partial v}{\partial y} \right) = -\frac{\partial p}{\partial y} + \mu \left( \frac{\partial^2 v}{\partial x^2} + \frac{\partial^2 v}{\partial y^2} \right) + \frac{1}{3} \mu \frac{\partial}{\partial y} \left( \frac{\partial u}{\partial x} + \frac{\partial v}{\partial y} \right). \tag{S15}$$

The equations above can be scaled with the same parameters as previously.

Scaling of the first equation:

$$\frac{RT_0}{M}\rho_0\rho^*\left(\frac{1}{L_x}u^*\frac{\partial u^*}{\partial x^*}+\frac{1}{L_y}v^*\frac{\partial v^*}{\partial y^*}\right) \quad (S16)$$

$$=-\frac{p_0}{L_x}\frac{\partial p}{\partial x}+\mu\sqrt{\frac{RT_0}{M}}\left(\frac{1}{L_x^2}\frac{\partial^2 u^*}{\partial(x^*)^2}+\frac{1}{L_y^2}\frac{\partial^2 u^*}{\partial(y)^2}\right)$$

$$+\frac{1}{3}\mu\sqrt{\frac{RT_0}{M}}\left(\frac{1}{L_x^2}\frac{\partial^2 u^*}{\partial(x^*)^2}+\frac{1}{L_xL_y}\frac{\partial^2 u^*}{\partial x^*\partial y^*}\right).$$

By multiplying both sides with $\frac{L_x}{p_0}$ we get

$$\frac{L_xRT_0}{Mp_0}\rho_0\rho^*\left(\frac{1}{L_x}u^*\frac{\partial u^*}{\partial x^*}+\frac{1}{L_y}v^*\frac{\partial v^*}{\partial y^*}\right) \quad (S17)$$

$$=-\frac{\partial p^*}{\partial x^*}+\frac{\mu L_x}{p_0}\sqrt{\frac{RT_0}{M}}\left(\frac{1}{L_x^2}\frac{\partial^2 u^*}{\partial(x^*)^2}+\frac{1}{L_y^2}\frac{\partial^2 u^*}{\partial(y^*)^2}\right)$$

$$+\frac{1}{3}\frac{\mu L_x}{p_0}\sqrt{\frac{RT_0}{M}}\left(\frac{1}{L_x^2}\frac{\partial^2 u^*}{\partial(x^*)^2}+\frac{1}{L_xL_y}\frac{\partial^2 u^*}{\partial x^*\partial y^*}\right).$$

Let us notice that

$$\frac{RT_0}{Mp_0}=\frac{1}{\rho_0}.$$

Then the final form of the scaled equation is

$$\rho^*\left(u^*\frac{\partial u^*}{\partial x^*}+\frac{L_x}{L_y}v^*\frac{\partial v^*}{\partial y^*}\right) \quad (S18)$$

$$=-\frac{\partial p^*}{\partial x^*}+\frac{\mu}{L_xp_0}\sqrt{\frac{RT_0}{M}}\frac{\partial^2 u^*}{\partial(x^*)^2}+\frac{\mu L_x}{p_0L_y^2}\sqrt{\frac{RT_0}{M}}\frac{\partial^2 u^*}{\partial(y^*)^2}$$

$$+\frac{1}{3}\frac{\mu}{L_xp_0}\sqrt{\frac{RT_0}{M}}\frac{\partial^2 u^*}{\partial(x^*)^2}+\frac{1}{3}\frac{\mu}{L_yp_0}\sqrt{\frac{RT_0}{M}}\frac{\partial^2 u^*}{\partial x^*\partial y^*}.$$

One can see that all the four red marked terms on the RHS of the equation above can be reduced to

$$\frac{\mu}{L_xp_0}\sqrt{\frac{RT_0}{M}}$$

by properly multiplying with the non-dimensional number $\frac{L_x}{L_y}$.

3. Energy balance equation in 2D compressible flow

$$\rho c_p \left( u \frac{\partial T}{\partial x} + v \frac{\partial T}{\partial y} \right) \tag{S19}$$

$$= \frac{\partial}{\partial x}\left(k\frac{\partial T}{\partial x}\right) + \frac{\partial}{\partial y}\left(k\frac{\partial T}{\partial y}\right) + \beta T \left( u \frac{\partial p}{\partial x} + v \frac{\partial p}{\partial y} \right)$$
$$+ 2\mu\left[\left(\frac{\partial u}{\partial x}\right)^2 + \left(\frac{\partial v}{\partial y}\right)^2\right] + \mu\left(\frac{\partial u}{\partial y} + \frac{\partial v}{\partial x}\right)^2$$
$$- \frac{2}{3}\mu\left(\frac{\partial u}{\partial x} + \frac{\partial v}{\partial y}\right)^2.$$

$T_0$ is used here as the scaling parameter for temperature $\left(T^* = \frac{T}{T_0}\right)$ and thermal expansion coefficient $(\beta^* = \beta T_0)$.

Scaling the equation as previously gives the following expression:

$$c_p \rho_0 T_0 \sqrt{\frac{RT_0}{M}} \rho^* \left( \frac{1}{L_x} u^* \frac{\partial T^*}{\partial x^*} + \frac{1}{L_y} v^* \frac{\partial T^*}{\partial y^*} \right) \tag{S20}$$
$$= kT_0 \left( \frac{1}{L_x^2} \frac{\partial^2 T^*}{\partial (x^*)^2} + \frac{1}{L_y^2} \frac{\partial^2 T^*}{\partial (y^*)^2} \right)$$
$$+ p_0 \sqrt{\frac{RT_0}{M}} \beta^* T^* \left( \frac{1}{L_x} u^* \frac{\partial p^*}{\partial x^*} + \frac{1}{L_y} v^* \frac{\partial p^*}{\partial y^*} \right)$$
$$+ 2\mu \frac{RT_0}{M} \left[ \frac{1}{L_x^2} \left(\frac{\partial u^*}{\partial x^*}\right)^2 + \frac{1}{L_y^2} \left(\frac{\partial v^*}{\partial y^*}\right)^2 \right]$$
$$+ \mu \frac{RT_0}{M} \left( \frac{1}{L_x} \frac{\partial u^*}{\partial y^*} + \frac{1}{L_y} \frac{\partial v^*}{\partial x^*} \right)^2$$
$$- \frac{2}{3} \mu \frac{RT_0}{M} \left( \frac{1}{L_x} \frac{\partial u^*}{\partial x^*} + \frac{1}{L_y} \frac{\partial v^*}{\partial y^*} \right)^2.$$

After expanding all the terms and dividing by $\mu \frac{RT_0}{ML_x^2}$ we get:

$$\frac{c_p \rho_0 T_0 M L_x}{\mu RT_0} \sqrt{\frac{RT_0}{M}} \rho^* u^* \frac{\partial T^*}{\partial x^*} + \frac{c_p \rho_0 T_0 M L_x^2}{\mu RT_0 L_y} \sqrt{\frac{RT_0}{M}} \rho^* v^* \frac{\partial T^*}{\partial y^*} \tag{S21}$$
$$= \frac{kM}{\mu R} \frac{\partial^2 T^*}{\partial (x^*)^2} + \frac{kML_x^2}{\mu R L_y^2} \frac{\partial^2 T^*}{\partial (y^*)^2} + \frac{ML_x p_0}{\mu RT_0} \sqrt{\frac{RT_0}{M}} \beta^* T^* u^* \frac{\partial p^*}{\partial x^*}$$
$$+ \frac{ML_x^2 p_0}{\mu RT_0 L_y} \sqrt{\frac{RT_0}{M}} \beta^* T^* v^* \frac{\partial p^*}{\partial y^*} + 2\left(\frac{\partial u^*}{\partial x^*}\right)^2 + 2\frac{L_x^2}{L_y^2}\left(\frac{\partial v^*}{\partial y^*}\right)^2$$
$$+ \left(\frac{\partial u^*}{\partial y^*}\right)^2 + 2\frac{L_x}{L_y}\frac{\partial u^*}{\partial y^*}\frac{\partial v^*}{\partial x^*} + \frac{L_x^2}{L_y^2}\left(\frac{\partial v^*}{\partial x^*}\right)^2 - \frac{2}{3}\left(\frac{\partial u^*}{\partial x^*}\right)^2$$
$$- \frac{4}{3}\frac{L_x}{L_y}\frac{\partial u^*}{\partial x^*}\frac{\partial v^*}{\partial y^*} - \frac{2}{3}\frac{L_x^2}{L_y^2}\left(\frac{\partial v^*}{\partial y^*}\right)^2.$$

One can see that the terms $\frac{c_p \rho_0 T_0 M L_x^2}{\mu RT_0 L_y} \sqrt{\frac{RT_0}{M}}$ can be reduced to $\frac{c_p \rho_0 T_0 M L_x}{\mu RT_0} \sqrt{\frac{RT_0}{M}}$ when multiplied by a dimensionless number $\frac{L_y}{L_x}$. Substituting $c_p = \frac{5R}{2M}$ for ideal gas, allows us to further convert it to the term present on the RHS of the equation: $\frac{5ML_xp_0}{2\mu RT_0}\sqrt{\frac{RT_0}{M}}$. Let us note that this expression is the reciprocal of the non-dimensional number derived from the momentum equation:

$$\frac{1}{\frac{\mu}{L_xp_0}\sqrt{\frac{RT_0}{M}}} = \frac{ML_xp_0}{\mu RT_0}\sqrt{\frac{RT_0}{M}} = \frac{L_xp_0}{\mu}\sqrt{\frac{M}{RT_0}}$$

Similarly $\frac{kML_x^2}{\mu RL_y^2}$ is reduced to $\frac{kM}{\mu R} = \frac{5}{2}\frac{k}{\mu c_p}$ which corresponds to reversed Prandtl number (with $\frac{5}{2}$ constant) for an ideal gas.

The final parameters of the internal energy function obtained in the dimensionless analysis are:

1) $\frac{L_y}{L_x}$
2) $\frac{L_xp_0}{\mu}\sqrt{\frac{M}{RT_0}}$
3) $Pr = \frac{\mu c_p}{k}$